\documentclass[aps,pre,twocolumn,showpacs,10pt,superscriptaddress]{revtex4-1}

\usepackage{graphicx}% Include figure files
\usepackage{dcolumn}% Align table columns on decimal point
\usepackage{bm}% bold math
\usepackage{amsmath}
\usepackage{xcolor}
\usepackage{amssymb}
\usepackage{float}
\usepackage[titletoc,title]{appendix}
\usepackage{lipsum}

\newcommand{\bra}[1]{\langle #1 |}
\newcommand{\ket}[1]{| #1 \rangle}
\begin{document}

%%%%%%%%%%%%%%%%%%%%%%%%%%%%%%%%%%%%%%%%%%%%%%%%%%%%%%%%%%%%%%

\author{Supanut Thanasilp}
\email{supanut.thanasilp@u.nus.edu}
\affiliation{Centre for Quantum Technologies, National University of Singapore, 3 Science Drive 2, Singapore 117543}
\author{Jirawat Tangpanitanon}
\affiliation{Centre for Quantum Technologies, National University of Singapore, 3 Science Drive 2, Singapore 117543}
\author{Marc-Antoine Lemonde}
\affiliation{Centre for Quantum Technologies, National University of Singapore, 3 Science Drive 2, Singapore 117543}
\author{Ninnat Dangniam}
\affiliation{Department of Physics and Center for Field Theory and Particle Physics, Fudan University, Shanghai 200433, China}
\author{Dimitris G. Angelakis}
\email{dimitris.angelakis@qubit.org}
\affiliation{Centre for Quantum Technologies, National University of Singapore, 3 Science Drive 2, Singapore 117543}
\affiliation{School of Electrical and Computer Engineering, Technical University of Crete, Chania, Greece 73100}

%%%%%%%%%%%%%%%%%%%%%%%%%%%%%%%%%%%%%%%%%%%%%%%%%%%%%%%%%%%%%

\title{Quantum supremacy and quantum phase transitions}

%Quantum advantage with analog quantum processors for \\ material science and machine learning}

\date{\today}
%%%%%%%%%%%%%%%%%%%%%%%%%%%%%%%%%%%%%%%%%%%%%%%%%%%%%%%%%%%%%

\begin{abstract}
Demonstrating the ability of existing quantum platforms to perform certain computational tasks intractable to classical computers represents a cornerstone in quantum computing.
Despite the growing number of such proposed ``quantum supreme'' tasks, it remains an important challenge to identify their direct applications.
In this work, we describe how the approach proposed in Ref.~[arXiv:2002.11946] for demonstrating quantum supremacy in generic driven analog many-body systems, such as those found in cold atom and ion setups,  can be extended to explore dynamical quantum phase transitions.
We show how key quantum supremacy signatures, such as the distance between the output distribution and the expected Porter Thomas distribution at the supremacy regime, can be used as effective order parameters.
We apply this approach to a periodically driven disordered 1D Ising model and show that we can accurately capture the transition between the driven thermalized and many-body localized phases. 
This approach also captures the transition towards the Floquet prethermalized regime for high-frequency driving. 
Revisiting quantum phases of matter under the light of the recent discussions about quantum supremacy draws a link between complexity theory and analog many-body systems.
\end{abstract}

\maketitle

%%%%%%%%%%%%%%%%%%%%%%%%%%%%%%%%%%%%%%%%%%%%%%%%%%%%%%%%%%%%%%%%%%%%%%%%%%%%%%%%%%%%%
%%%%%%%%%%%%%%%%%%%%%%%%%%%%%%%%%%%%%%%%%%%%%%%%%%%%%%%%%%%%%%%%%%%%%%%%%%%%%%%%%%%%%

\section{Introduction}
Recent experimental developments~\cite{2019_martinis_nat,2018_neill_sci,PhysRevLett.123.250503,bs_supremacy,bs_review} have made the first steps toward demonstrating the ability of noisy intermediate-scale quantum (NISQ) devices~\cite{2018_preskill_quantum} to perform certain computational tasks intractable to classical computers~\cite{2017_Harrow_Nat,Ralph, 2011_shepherd,2018_hartmut_natphy, Umesh,qrc_Movassagh:2019eog,2011_Aaronson_PTC, Aaronson2011,bs_review,2018_eisert_prx,PhysRevLett.118.040502,born_supremacy}. 
Early proposals to demonstrate this so-called quantum supremacy include sampling from random superconducting circuits~\cite{2018_hartmut_natphy, Umesh,qrc_movassagh2020} and boson sampling on photonic platforms~\cite{2011_Aaronson_PTC, Aaronson2011,bs_review}. Despite being an important milestone for quantum computation in the NISQ era, extending the proposed computational tasks to address real-world problems remains an open challenge. 
This is particularly true in the context of sampling from quantum random circuits where any practical applications remains elusive.
For example, the randomness required in such circuits makes them in principle unsuitable for medium- to large-scale variational quantum algorithms~\cite{2018_hartmut_natcom}, thus greatly limiting their applications in near-term quantum chemistry~\cite{chem_rev1,chem_rev2}, material sciences~\cite{chem_rev3} and quantum machine learning~\cite{qml_rev}. 

More recently, the computational task of sampling from generic closed periodically-driven quantum many-body systems has also been proposed to demonstrate quantum supremacy~\cite{P1}. This opens the road for a plethora of analog based quantum simulators on periodically driven cold atoms and ions to be used for such tasks~\cite{cold_rev, trapped_rev_monroe2020programmable}. The computational complexity of sampling from such dynamics is intimately related to the ability of the closed system to thermalize under the combined influence of interactions and the external drive.
Once thermalised, the associated temperature is effectively infinite resulting in a quantum evolution capable of exploring its entire Hilbert space~\cite{2014_Rigol_PRX, pre_review,mbl_driven_annals_ponte_2015,mbl_driven_2013_Alessio_AoP}.
This is in contrast with the driven many-body localized (MBL) phase where the presence of disorder prevents the system to thermalize and limits the quantum evolution to a restricted portion of the Hilbert space~\cite{mbl_driven_2013_Alessio_AoP,mbl_driven_annals_ponte_2015,mbl_driven_2015_Ponte_PRL,mbl_driven_2015_prl_Roderich,mbl_driven_2016_Abanin_AoP,mbl_Rev}.
Efficient theoretical representations exploiting this fact suggests that the sampling from MBL dynamics is tractable to classical computers~\cite{mbl_mps_jens_PhysRevLett,mbl_mps_spectral_PhysRevB.92.024201,mbl_mps_simon_prx_UMPO}. 
Therefore, the experiment we proposed in Ref.~\cite{P1} to achieve quantum supremacy can be interpreted as a protocol to probe how much of the Hilbert space has been explored under specific driven quantum dynamics.   
This ability of tuning in and out of the chaotic evolution underlying the thermal phase by controlling the level of disorder has already been exploited as a potential application in the context of quantum machine learning~\cite{qml_p3}.

In this work, we describe how the quantum supremacy experiment proposed in Ref.~\cite{P1} can be extended to probe the phase diagram of periodically-driven disordered many-body systems. More specifically, we propose utilising key quantum supremacy signatures as order parameters, such as the Kullback-Leibler divergence (KLD)~\cite{kld_1951} between the output probability distribution and the Porter-Thomas (PT) distribution~\cite{1956_pt}.
The one to one matching in the behaviour of these two distributions is a strong indication that the system is capable of exploring its entire Hilbert space~\cite{2018_hartmut_natphy} and thus intimately related to the computational complexity of the sampling task
%%%
\footnote{This is similar to the growth of the bipartite entanglement entropy in closed systems which also captures the increase in the capacity of the system to explore its Hilbert space~\cite{2016_rigol_ap}. 
While a large entanglement entropy implies a classically intractable matrix product state (MPS) representation~\cite{entropy_mps_sim_Cirac_Prl2008}, a vanishing KLD from the PT distribution has been shown to be a key condition to prove that {\it any} classical approach is intractable~\cite{2018_hartmut_natphy,P1}.}.
%%%
Therefore, such order parameter offers additional insights into the dynamics compared to the measure of the population imbalance commonly used in experiments~\cite{mbl_Rev,mbl_exp_2019_prx_bloch_2D,mbl_exp_2019_prl_bloch,mbl_exp_scq_chiaro2019direct}.
In addition, the measurement of the KLD from the PT distribution, performed only in the computational basis, should also be experimentally more accessible compared to probing the level-statistics~\cite{2016_rigol_ap} which is already challenging for undriven systems~\cite{mbl_exp_2017_Roushan_Sci} with no known efficient extension to the driven cases.
We show that probing the KLD from the PT distribution allows to distinguish the driven thermalized phase from the driven MBL phase as well as from the Floquet prethermalized regime~\cite{pre_review,pre_exp_prx2020_Bloch}. 
The latter corresponds to the scenario where the drive frequency exceeds all relevant energy scales of the system, leading to long-lived prethermalization dynamics. 
Although we focus on the KLD between the output and the PT distributions as it is a straightforward quantum supremacy signature, we emphasize that our work could be extended to any suitable metrics associated with the computational complexity of sampling from the different phases. For example, we discuss in Appendix~\ref{App:EE} how the entanglement entropy over multiple sub-systems can also be used as an alternative supremacy signature for probing the phase diagram.

As a specific model, we focus on an isolated Ising spin chain with nearest-neighbour interactions, disordered on-site energies and a periodically time-varying global magnetic field. 
We analyse the dynamics using the Floquet formalism which allows to highlight the role played by the effective long-range multi-body interactions generated by the drive in the context of quantum supremacy.
Our work thus proposes a direct application of a quantum supremacy experiment and present additional physical intuitions about the origin of the computational complexity associated with sampling from a driven thermalized quantum system.

%%%%%%%%%%%%%%%%%%%%%%%%%%%%%%%%%%%%%%%%%%%%%%%%%%%%%%%%%%%%%%%%%%%%%%%%%%%%%%%%%%%%%
%%%%%%%%%%%%%%%%%%%%%%%%%%%%%%%%%%%%%%%%%%%%%%%%%%%%%%%%%%%%%%%%%%%%%%%%%%%%%%%%%%%%%

\begin{figure}
	\includegraphics[width=0.95\columnwidth]{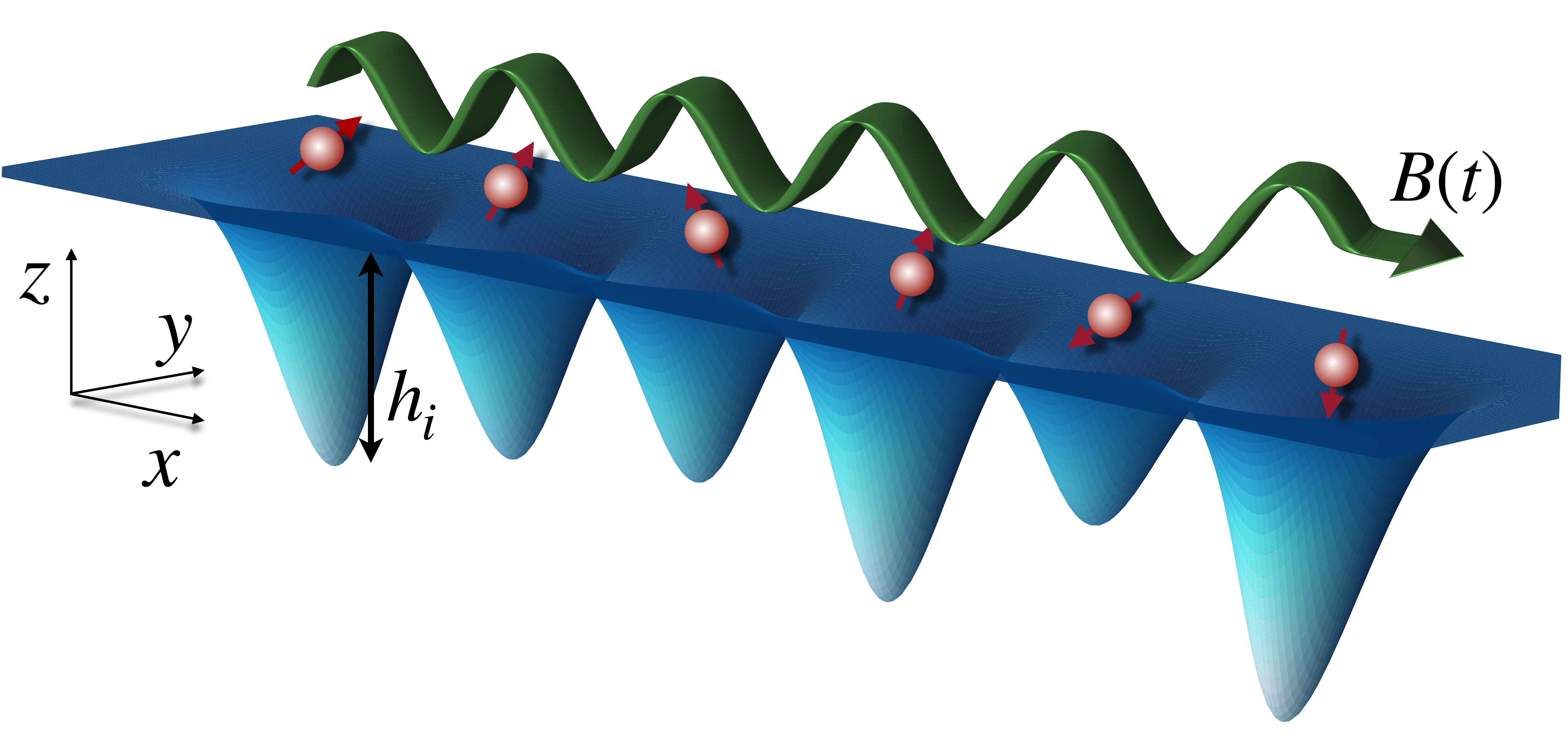}
	\caption{Schematic of a periodically driven quantum Ising chain with nearest-neighbor interactions and disordered onsite energies.} %$h_i \in [-W/2, W/2]$.}
	\label{fig0}
\end{figure}

%%%%%%%%%%%%%%%%%%%%%%%%%%%%%%%%%%%%%%%%%%%%%%%%%%%%%%%%%%%%%%%%%%%%%%%%%%%%%%%%%%%%%
%%%%%%%%%%%%%%%%%%%%%%%%%%%%%%%%%%%%%%%%%%%%%%%%%%%%%%%%%%%%%%%%%%%%%%%%%%%%%%%%%%%%%

\section{Quantum phases in driven disordered Ising spin chains}

\subsection{Model}

The particular model we consider is sketched in Fig.~\ref{fig0} and consists of a periodically-driven Ising chain with disordered onsite energies. The corresponding Hamiltonian reads $\hat{H}(t) = \hat{H}_0 + \hat{H}_d(t)$, where
\begin{gather}\label{H_ising}
\hat{H}_0 =\sum_{i=1}^L h_i \hat{\sigma}^z_i + B_0\sum_{i=1}^L\hat{\sigma}_i^x + J\sum_{i=1}^{L-1}\hat{\sigma}^z_i\hat{\sigma}^z_{i+1}, \\ 
\hat{H}_d(t) = \delta B \cos{(\omega t)}\sum_{i=1}^L\hat{\sigma}^x_i.
\end{gather}
Here $\hat{\sigma}_i^x$ ($\hat{\sigma}_i^z$) is the $x$ ($z$) Pauli matrix acting on the $i^{\rm th}$ spin of an open chain of $L$ total sites, $J$ represents the nearest-neighbour interaction strength and $B(t) = B_0 + \delta B\cos(\omega t)$ represents a periodic magnetic field along the $x$-axis of frequency $\omega \equiv 2\pi/T$. 
We consider disordered static magnetic fields along the $z$-axis with strengths $h_i \in \left[-W/2,W/2\right]$ being drawn from a uniform distribution ranging from $-W/2$ to $W/2$. 
Importantly, for finite disorder $W\neq0$ the Hamiltonian at different times $t_1$ and $t_2 \neq t_1 + nT$ with $n \in \mathbb{Z}$ do not commute, i.e.~$[\hat H(t_1),\hat H(t_2)] \neq 0$, and the contribution from the drive averages to zero over a period, i.e.~$\frac{1}{T}\int_0^T \hat{H}_d(t) = 0$. 

The time evolution over a driving period is described by the unitary operator
\begin{equation} \label{exact_U}
\hat{U}=\hat{\mathcal{T}}\exp \left[-i\int_0^T \hat{H}(t) dt\right] \equiv \exp \left[-i \hat{H}_F T\right],
\end{equation}
where $\hat{\mathcal{T}}$ is the time-ordering operator. The time-independent Hamiltonian  $\hat{H}_F$ is known as the Floquet Hamiltonian and fully describes the dynamics at stroboscopic times $t_n = nT$.
While a specific form of the Ising model is studied in this work, the results obtained can be generalized for other driven many-body systems such as Bose-Hubbard model or different variations of the Ising spin model. All together, such driven systems have been experimentally implemented in various quantum platforms including superconducting circuits~\cite{2018_neill_sci}, cold atoms~\cite{mbl_exp_driven_2017_bloch_natphy,pre_exp_prx2020_Bloch,cold_rev} and trapped ions~\cite{trapped_rev_monroe2020programmable}.

%Such driven disordered models have been implemented in different quantum platforms~\cite{mbl_exp_driven_2017_bloch_natphy,2018_neill_sci}.

%%%%%%%%%%%%%%%%%%%%%%%%%%%%%%%%%%%%%%%%%%%%%%%%%%%%%%%%%%%%%%%%%%%%%%%%%%%%%%%%%%%%%
%%%%%%%%%%%%%%%%%%%%%%%%%%%%%%%%%%%%%%%%%%%%%%%%%%%%%%%%%%%%%%%%%%%%%%%%%%%%%%%%%%%%%

\subsection{Phases and level-statistics}

\begin{figure*}
	\includegraphics[width=1\textwidth]{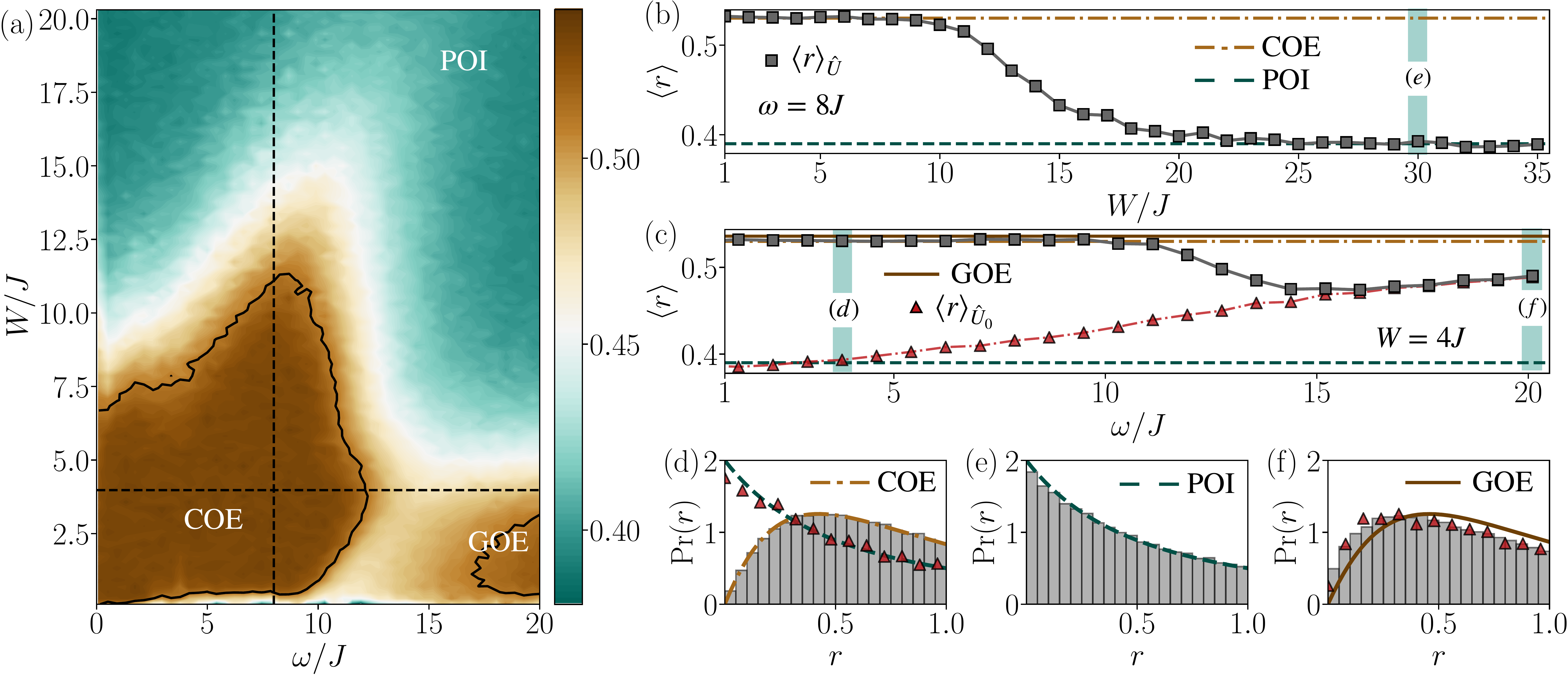}
	\caption{\textbf{Level statistics of the driven disordered Ising spin chain.} (a) Average level spacing $\langle r \rangle_{\hat{U}}$ of the full unitary $\hat U$ as a function of the disorder strength $W/J$ and the driving frequency $\omega / J$ for $L=9$ and $B_0 = -\delta B = 1.25 J$.
	The three limiting statistics are indicated in white and the solid black line represents $\langle r \rangle_{\hat U} = 0.51$. 
	(b) A cut along the frequency $\omega = 8J$ [vertical dashed line in (a)] shows $\langle r \rangle_{\hat U}$ as a function of $W/J$ and captures the phase transition from driven thermalized to the driven MBL. (c) A cut along $W = 4J$ [horizontal dashed line in (a)] captures the transition as a function of $\omega/J$ from the driven thermalized phase to the prethermalized regime. The red triangles represent the mean value $\langle r \rangle_{\hat U_0}$ computed only considering the evolution under $\hat H_0$ for a time $t = 2\pi/\omega$.
	(d)-(f) Full distribution ${\rm Pr}(r)$ for the driven thermalized phase ($W = 4J, \omega = 4.2J $) following the COE, MBL phase ($W = 30 J, \omega = 8J $) following POI and the prethermalization regime ($W = 4J, \omega = 20.1J $) converging towards the GOE respectively. 
	In (d), the statistics of $\hat U$ and $\hat U_0$ drastically differ while they converge in (f); contrasting the importance of the the multi-body long-range interaction induced by the drive at different $\omega$. Each data point results from $100$ disorder realizations.}
	\label{fig2}
\end{figure*}

Depending on the disorder strength $W/J$ and driving frequency $\omega/J$, the system can be in three distinct regimes~\cite{pre_review,mbl_Rev}. 
At low disorder strengths, the system thermalizes under its own dynamics if given enough time and follows the eigenstate thermalization hypothesis~\cite{pre_review}.
Under the assumption that the system can be considered closed over the entire process, the constant energy input provided by the drive leads to an effective temperature that is infinite~\cite{2014_Rigol_PRX, pre_review,mbl_driven_annals_ponte_2015,mbl_driven_2013_Alessio_AoP}. 
The time-scale over which this thermalization process takes place strongly depends on the drive frequency~\cite{pre_review,pre_exp_prx2020_Bloch}. 
For low-frequency drives, the system can efficiently respond and rapidly reaches this infinite-temperature limit corresponding to what is known as the {\it driven thermalized phase}~\cite{2014_Rigol_PRX, pre_review,mbl_driven_annals_ponte_2015,mbl_driven_2013_Alessio_AoP}. 
In the case where the driving frequency exceeds all relevant energy scales of the system however, the time required for thermalizing can be greatly extended leading to a long-lived {\it prethermalization regime}~\cite{pre_review,pre_exp_prx2020_Bloch}.
Finally, in presence of large disorders, the system fails to thermalize at any time and is said to be in the {\it driven MBL phase}~\cite{mbl_driven_2015_Ponte_PRL,mbl_driven_2016_Abanin_AoP,mbl_driven_2013_Alessio_AoP,mbl_driven_2015_prl_Roderich,mbl_driven_annals_ponte_2015}.
The two phases and the prethermalized regime are depicted in Fig.~\ref{fig2} (a). 

%%%%%%%%%%%%%%%%%%%%%%%%%%%%%%%%%%%%%%%%%%%%%%%%%%%%%%%%%%%%%%%%%%%%%%%%%%%%%%%%%%%%%
%%%%%%%%%%%%%%%%%%%%%%%%%%%%%%%%%%%%%%%%%%%%%%%%%%%%%%%%%%%%%%%%%%%%%%%%%%%%%%%%%%%%%
%%%%%%%%%%%%%%%%%%%%%%%%%%%%%%%%%%%%%%%%%%%%%%%%%%%%%%%%%%%%%%%%%%%%%%%%%%%%%%%%%%%%%
%%%%%%%%%%%%%%%%%%%%%%%%%%%%%%%%%%%%%%%%%%%%%%%%%%%%%%%%%%%%%%%%%%%%%%%%%%%%%%%%%%%%%

One of the standard approach to distinguish the different phases is based on the notion of level statistics of the unitary operator $\hat{U}$~\cite{2014_Rigol_PRX,mbl_driven_2015_Ponte_PRL}. 
Let $|\phi_n\rangle$ be an eigenstate of the Floquet Hamiltonian with eigenvalue $\epsilon_n$, i.e.~$\hat{H}_F|\phi_n\rangle = \epsilon_n|\phi_n\rangle$, it follows that  
\begin{equation}
\hat{U}=\sum_n e^{i\theta_n}|\phi_n\rangle\langle \phi_n|,
\end{equation}
where $\theta_n = \epsilon_n T \text{ modulo 2}\pi$. 
The level statistics is defined as the normalized distribution $\text{Pr}(r)$ of the level spacing 
\begin{equation}
r_n= \frac{\text{min}(\delta_n,\delta_{n+1})}{\text{max}(\delta_n,\delta_{n+1})},
\end{equation}
with $\delta_n = \theta_{n+1}-\theta_n$ and $\theta_{n+1} \geq \theta_n$.

In the driven thermalized phase, the level statistics of $\hat U$ coincides with the circular orthogonal ensemble (COE) statistics, given by
\begin{align}
\text{Pr}_{\rm COE}(r)=&\frac{2}{3}\left[\frac{\sin\left(\frac{2\pi r}{r+1}\right)}{2\pi r^2}+\frac{1}{(r+1)^2}+\frac{\sin\left(\frac{2\pi}{r+1}\right)}{2\pi} \right.\nonumber \\ &\left.-\frac{\cos\left(\frac{2\pi}{r+1}\right)}{r+1}-\frac{\cos\left(\frac{2\pi r}{r+1}\right)}{r(r+1)}\right].
\end{align}
This result means that, from a statistical point of view, a generic $\hat{U}$ associated with this phase is equivalent to a unitary matrix randomly sampled from the complete ensemble of unitaries that conserves the time-reversal symmetry~\cite{2010_haake}. 
The fact that Pr$_{\rm COE}(0) = 0$ indicates phase repulsion and thus correlations between the different eigenstates. 
A COE evolution in low-frequency driven systems is an indicator of thermalization to infinite temperatures~\cite{2014_Rigol_PRX}.

The scenario is completely different for systems in the driven MBL phase where the level statistics is described by a Poisson distribution~\cite{mbl_driven_2016_Abanin_AoP},
\begin{equation}
\text{Pr}_{\rm POI}(r) = \frac{2}{(1+r)^2}.
\end{equation}
This time, $r=0$ corresponds to the peak of $\text{Pr}_{\rm POI}(r)$ indicating a high level of degeneracies in the MBL phase. This absence of phase repulsion is intuitive as spatially distant localized eigenstates are uncorrelated. 

In the limit where the driving frequency exceeds all relevant energy scales of the system, a slightly different interpretation of the level statistics is required. This comes as all phases $\theta_n = \epsilon_n T < 2\pi$ (for $\omega > {\rm max}[\epsilon_n]$) become strictly linearly related to the eigenenergies of $\hat H_F$ so that the statistics of $\hat U$ must coincide with the statistics of $\hat H_F$. 
In addition, for $\omega \rightarrow \infty$, $\hat H_F \rightarrow \hat H_0$ (see section \ref{Sec:Magnus}) and one should expect to recover the statistics of the undriven Ising model as the frequency increases. 
Similar to the driven case, $\hat H_0$ has two distinct phases: the MBL phase for large disorder and the {\it finite-temperature} thermalized phase for weak disorder. 
%This time however, the effective temperature in undriven thermalized systems is finite. 
While the undriven MBL still leads to the Poisson statistics, $\text{Pr}(r)$ of the undriven thermalized phase follows the Gaussian orthogonal ensemble (GOE),
\begin{align}
\text{Pr}_{\rm GOE}(r) = \frac{27}{4}\frac{r+r^2}{(1+r+r^2)^{5/2}},
\end{align}
which corresponds to the ensemble of matrices with independent normal random variables as elements and subjected to the orthogonality constraint~\cite{2010_haake}.
This high frequency limit is known as the prethermalization regime and
we note that in the thermodynamics limit, the energy spectrum of the system becomes infinite so that the driving frequency required to reach this regime also becomes infinite.

In Fig.~\ref{fig2}, we show the level statistics as a function of the drive frequency $\omega/J$ and the disorder strength $W/J$ for an Ising chain of $L=9$. The results are obtained by exactly diagonalizing $\hat U$ of Eq.~\eqref{exact_U} over 100 disorder realizations for each data point.
In panels (a)-(c), we show the mean value of level spacing $\langle r \rangle_{\hat{U}}$ as a metric to characterize the different phases and compare it to the values obtained in the COE ($\langle r \rangle_{\rm COE} \approx 0.527$), the POI ($\langle r \rangle_{\rm POI} \approx 0.386$) and the GOE ($\langle r \rangle_{\rm GOE} \approx 0.536$).
We distinguish the three distinct regimes of the driven dynamics and explicitly show the transition from driven thermalized to driven MBL phases as we increase $W/J$ with fixed $\omega/J$ in panel (b) and from driven thermalized phase to prethermalization regime as we increase $\omega/J$ with fixed $W/J$ in panel (c). 
In panels (d)-(f), we show Pr$(r)$ obtained from $\hat{U}$ in these three regimes where it coincides with their respective statistical ensembles. 
Finally, we also compute the level statistics of the unitary operator produced by only considering the undriven Hamiltonian $\hat H_0$, i.e.~$\hat U_0 = \exp{-i \hat H_0 T}$, and compare it to the exact results corresponding to the same frequency $\omega = 2\pi/T$. As expected, $\langle r \rangle_{\hat{U}}$ and $\langle r \rangle_{\hat{U}_0}$ converge as $\omega/J$ increases.
We note that in absence of disorder, $W\rightarrow 0$, the system evades thermalization for any driving frequencies. This comes as the dynamics described by Eq.~\eqref{H_ising} with $W=0$ is equivalent to a non-interacting fermionic system~\cite{Russomanno_2016}.

%%%%%%%%%%%%%%%%%%%%%%%%%%%%%%%%%%%%%%%%%%%%%%%%%%%%%%%%%%%%%%%%%%%%%%%%%%%%%%%%%%%%%
%%%%%%%%%%%%%%%%%%%%%%%%%%%%%%%%%%%%%%%%%%%%%%%%%%%%%%%%%%%%%%%%%%%%%%%%%%%%%%%%%%%%%

\section{Quantum supremacy in driven analog many-body systems}
In this section, we discuss the computational complexity of sampling from quantum states evolving under the dynamics associated with the different regimes discussed above. 
We begin by briefly reviewing the formal evidences presented in Ref.~\cite{P1} which suggests that such a task is classically intractable in the driven thermalized phase following the COE. 
We focus on the convergence of the distribution of the output probabilities towards the PT distribution as a key quantum supremacy signature which we  contrast in the MBL phase and the prethermalized regime.

%%%%%%%%%%%%%%%%%%%%%%%%%%%%%%%%%%%%%%%%%%%%%%%%%%%%%%%%%%%%%%%%%%%%%%%%%%%%%%%%%%%%%
%%%%%%%%%%%%%%%%%%%%%%%%%%%%%%%%%%%%%%%%%%%%%%%%%%%%%%%%%%%%%%%%%%%%%%%%%%%%%%%%%%%%%

\subsection{Quantum supremacy in the driven thermalized phase}

We define the exact probability of measuring the basis state $\ket{{\bf z}} = \otimes_i^L \ket{z_i}$ after $m$ driving cycles as $p_m({\bf z}) \equiv |\bra{{\bf z}} \psi_m\rangle|^2$, where $\hat\sigma_i^z\ket{z_i} = z_i\ket{z_i}$ is the eigenstate of the single-qubit Pauli $z$ matrix acting on site $i$ and where $\ket{ \psi_m} = (\hat U)^m\ket{\psi_0}$. The initial state $\ket{\psi_0}$ is assumed to be a product state.
The computational task considered here is {\it the sampling of output bitstrings ${\bf z}$ from $p_m({\bf z})$ up to (realistic) additive errors.} More precisely, it requires to produce a sampler $\mathcal{C}$ capable of providing bitstrings from a distribution $q(\textbf{z})$ that is {\it additively closed} to the exact distribution $p_m(\textbf{z})$, i.e.
\begin{equation}\label{add_err}
	\sum_\textbf{z} |p_m(\textbf{z})-q(\textbf{z})| \le \beta_0,
\end{equation}
where $\beta_0$ is a small positive constant. 
The sampler $\mathcal{C}$ can be either constructed by a quantum or a classical device. Quantum-mechanically, this can be achieved by performing projective measurements on an analog quantum simulator capable of reproducing the exact driven Hamiltonian. The difference between $q(\textbf{z})$ and $p_m(\textbf{z})$ can be kept small if the noise from the environment and imperfections in the controls remains sufficiently low. 

Producing this sampler $\mathcal{C}$ with a classical computer is however not as natural and strong analytical evidence suggests that, given that $\hat U$ follows the COE statistics, it is classically intractable~\cite{P1}. 
One of the key features that ensures this computational complexity is the fact that the associated quantum dynamics anti-concentrates~\cite{Hangleiter2018anticoncentration}. Mathematically, this means that the distribution of the output probabilities $p_m({\bf z})$ respects the condition
\begin{equation} \label{Eq:AC}
	\text{Pr}(p_m(\textbf{z})>\delta/N) \ge \gamma,
\end{equation}
for all possible $\textbf{z}$. Here $\delta$ and $\gamma$ are positive constants, $N = 2^L$ is the dimension of the Hilbert space, and the distribution is computed over all experimentally implementable unitaries $\{\hat{U}\}$. Equation \eqref{Eq:AC} ensures that most of the basis states have non-zero probability of being measured.   
Following the anti-concentration condition, one can show that if $\hat U$ follows the COE, the existence of any efficient classical sampler $\mathcal C$ would imply the collapse of the complexity hierarchy to the third level~\cite{Umesh}. 

In the case where the dynamics is such that the output state $\ket{\psi_m}$ has an equal probability of being anywhere in the Hilbert space, its output distribution Pr$(p_m({\bf z}))$ follows the PT distribution~\cite{2018_hartmut_natphy},
\begin{equation} \label{Eq:PTD}
 \text{PT}(p) = Ne^{-Np} \qquad {\rm for} \qquad N\gg1,
\end{equation}
and the dynamics satisfies the anti-concentration condition Eq.~\eqref{Eq:AC} with $\delta = 1$ and $\gamma = 1/e$. 
As a consequence, the convergence of the output distribution toward the PT distribution is a key signature of quantum supremacy~\cite{P1,2018_hartmut_natphy,2018_neill_sci}. The difference between these two distributions can be measured by the KLD~\cite{kld_1951}, defined as ($p = p_m({\bf z})$ for readability)
\begin{equation} \label{Eq:KLD}
\text{KLD}(\text{Pr}(p)\parallel \text{PT}(p))\equiv\sum_{p}\text{Pr}(p)\log \left( \frac{ \text{Pr}(p) }{ \text{PT}(p)}\right)\geq 0.
\end{equation}
The KLD is zero only when $\text{Pr}(p)=\text{PT}(p)$ for all $p$. 

\begin{figure}
	\includegraphics[width= 1 \columnwidth]{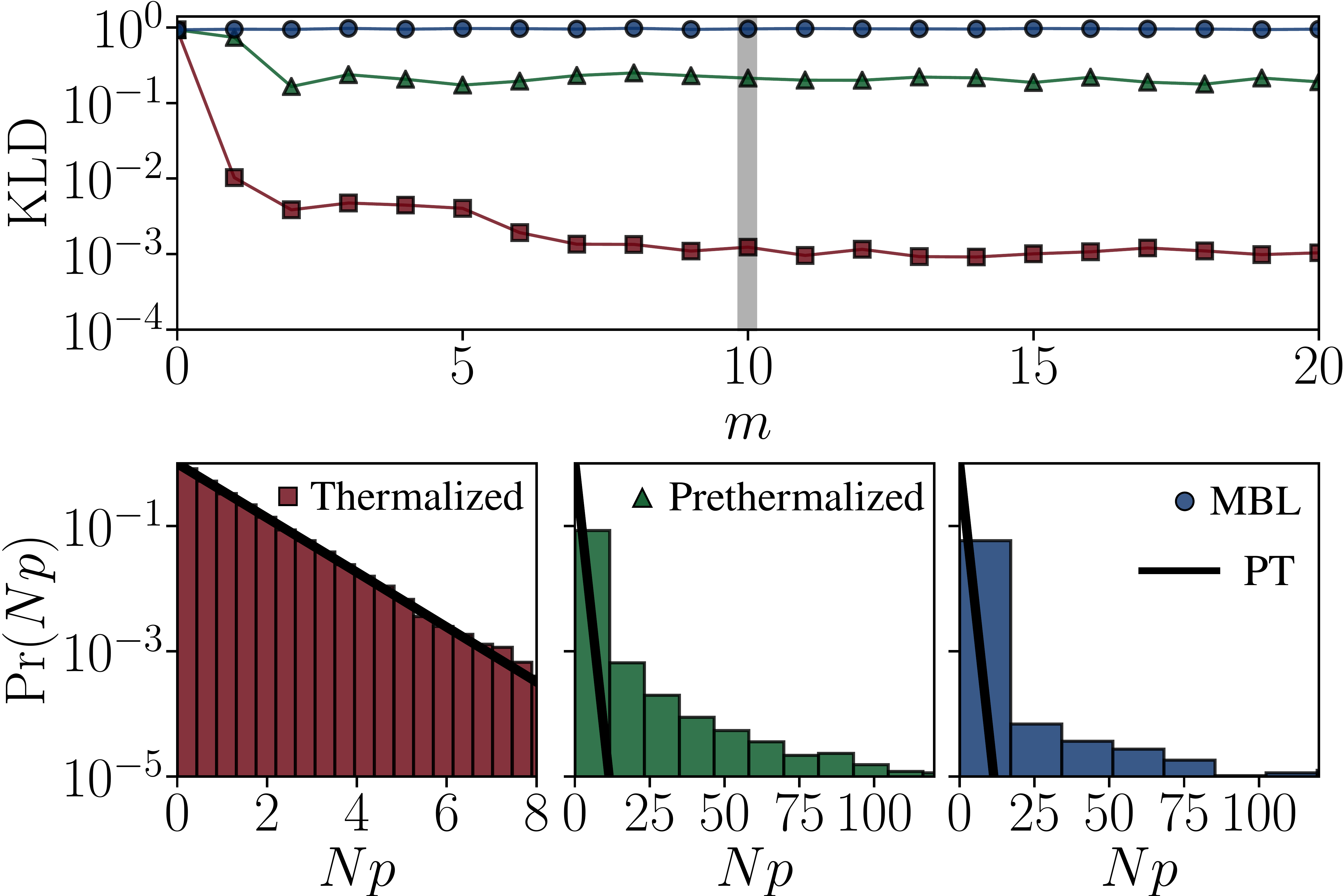}
	\caption{\textbf{Reaching the PT distribution.} (a)  $\text{KLD}(\text{Pr}(p)\parallel\text{PT}(p))$ with $p = p_m({\bf z})$ as a function of the driving cycle $m$ for the driven thermalized phase ($W=3J, \omega = 8J$; dark red squares) %and ($W=1.5J, \omega = 2J$; light red squares)
	, the driven MBL phase ($W=30J, \omega = 8J$; blue circles) and the prethermalization regime ($W=4J, \omega = 20 J$; green triangles). 
	(b)-(d) Output probability distributions ${\rm Pr}(Np)$ for $m=10$ in the three different regimes. (b) ${\rm Pr}(Np)$ in the driven thermalized phase ($\omega = 8J$) is in a good agreement with the PT distribution (black line) while it is not in (c) for the prethermalization regime and in (d) for the driven MBL phase. 
	The other parameters are $L=9$ with $N = 2^L$, $B_0 = -\delta B = 1.25 J$ and each data points has been computed over $100$ disorder realizations.}
	\label{fig3}
\end{figure}

In Fig.~\ref{fig3}, we plot the KLD between the output distribution Pr$(p=p_m({\bf z}))$ after $m$ driving cycles and the PT distribution PT$(p = p_m({\bf z}))$ for the three regimes of a driven Ising chain with $L=9$.
In panel (a), we show the evolution of the KLD as a function of $m$ and see that only in the driven thermalised phase Pr$(p)$ reaches the PT distribution after a number of cycles that increases as the driving frequency increases.

\subsection{Computational complexity of the MBL phase and the prethermalization regime} \label{Sec:MBL}

In the MBL phase, only a small fraction of the Hilbert space near the initial state is explored. 
Consequently, given an initial product state, $\ket{\psi_0} = \otimes_i^L \ket{\uparrow}$, only a limited number of output bitstrings can be measured after the quantum evolution and the anti-concentration condition can never be satisfied (see Appendix \ref{App:MBL} for more details). 
Consequently, the existence of a classical sampler $\mathcal C$ capable of providing bitstrings from a distribution additively closed to $p_m({\bf z})$ is not prohibited by the accepted hierarchy of the complexity classes.
%aforementioned quantum supremacy argument.

The more involved question of whether the simulation of the MBL dynamics is classically intractable or not has already been studied extensively~\cite{mbl_mps_jens_PhysRevLett,mbl_mps_spectral_PhysRevB.92.024201,mbl_mps_simon_prx_UMPO,mbl_Rev} and the  answer lies in the properties of the bipartite entanglement entropy. For an isolated system described by $\ket{\Psi}$, the bipartite entanglement entropy $S_e$ is obtained by dividing the system into two subsystems $\mathcal{S}$ and $\mathcal{B}$ and calculating the von Neumann entropy of one of the reduced density matrix, i.e.~$S_e = -{\rm Tr}_\mathcal{S} \{ \hat\rho_\mathcal{S}\log_2\hat\rho_\mathcal{S}\}$ with $\hat\rho_\mathcal{S} = {\rm Tr}_\mathcal{B}\{\ket\Psi\bra{\Psi}\}$.
In the MBL phase, all eigenstates of $\hat U$ obey the ``area law''~\cite{mbl_Rev}, meaning that $S_e$ is independent of the subsystem size in 1D systems. 
This property suggests an efficient classical representation using matrix product states (MPS) with low bond dimensions and the use of density matrix renormalization group (DMRG) techniques~\cite{mbl_mps_simon_prx_UMPO, dmrg_mps_review}. 
Moreover, given an initial product state, the entanglement entropy evolves logarithmically with time~\cite{mbl_ent_log_numeric,mbl_ent_log_theory}, allowing efficient and accurate classical simulation of long-time dynamics.  
Therefore, further supporting that driven MBL dynamics never converges towards the PT distribution which would represent infinite-temperature thermalization and would imply maximal entanglement entropy. In Appendix~\ref{App:MBL}, we analytically show that the MBL dynamics can be efficiently approximated with MPS up to additive errors.

The scenario in the prethermalized regime where $\hat U$ tends towards the GOE statistics is very different and many questions remain open. For example, the entanglement entropy of the eigenstates of $\hat U_{\rm GOE}$ obeys the ``volume law'' and an initial product state would see its entanglement entropy increase linearly in time~\cite{2016_Alessio_AiP}. Both properties force an efficient MPS representation to fail. 
On the other hand, the fact that the output distribution does not follow the PT distribution leaves uncertain whether or not
the output distribution anti-concentrates [see Eq.~\eqref{Eq:AC}]. 
The complexity class associated with sampling from a quantum state in the prethermalized regime thus remains unclear. 

%a quantum state in this regime do not tend to follow the PT distribution due to the associated finite temperature is a necessary (but not sufficient) condition for its probability distribution to concentrate [cf.~Eq.~\eqref{Eq:AC}], which would invalidate the proof for quantum supremacy presented in Ref.~\cite{P1}.
%The complexity class associated with sampling from a quantum state in the prethermalized regime thus remains unclear. 

%%%%%%%%%%%%%%%%%%%%%%%%%%%%%%%%%%%%%%%%%%%%%%%%%%%%%%%%%%%%%%%%%%%%%%%%%%%%%%%%%%%%%
%%%%%%%%%%%%%%%%%%%%%%%%%%%%%%%%%%%%%%%%%%%%%%%%%%%%%%%%%%%%%%%%%%%%%%%%%%%%%%%%%%%%%

\begin{figure*}
\includegraphics[width=0.95\textwidth]{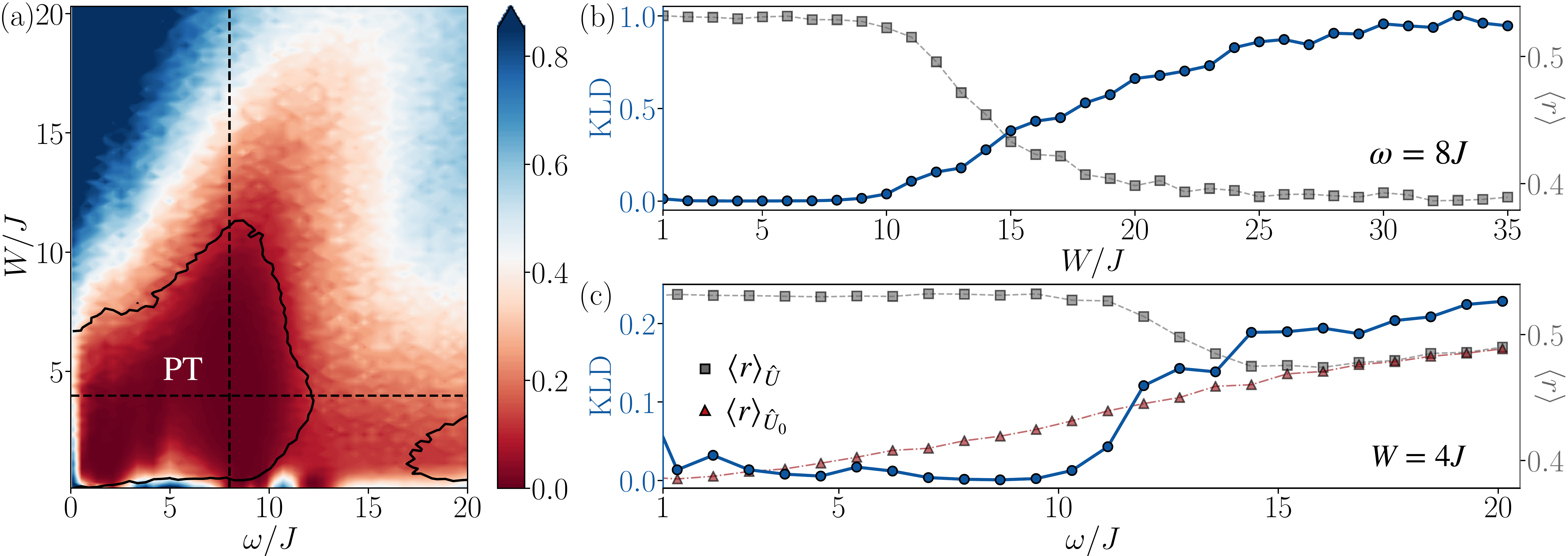}
\caption{\textbf{Probing the phase diagram with the $\text{KLD}(\text{Pr}(p)\parallel\text{PT}(p))$.} 
(a) $\text{KLD}(\text{Pr}(p)\parallel\text{PT}(p))$ with $p = p_m({\bf z})$ after $m=10$ driving cycles as a function of the disorder strength $W/J$ and the driving frequency $\omega / J$. 
The solid black line represents the parameters for which $\langle r \rangle_{\hat{U}} = 0.51$ as a comparison with Fig.~\ref{fig2} (a).
(b) A cut along the frequency $\omega = 8J$ (vertical dashed line in (a)) comparing the $\text{KLD}(\text{Pr}(p)\parallel\text{PT}(p))$ (blue dot, left y-axis) with $\langle r \rangle_{\hat U}$ (gray squares, right y-axis) as a function of $W/J$ along the phase transition from driven thermalized to driven MBL. 
(c) A cut along $W = 4J$ (horizontal dashed line in (a)) as a function of $\omega/J$ demonstrates the transition towards the prethermalization regime. 
This time, we compare the $\text{KLD}(\text{Pr}(p)\parallel\text{PT}(p))$ with $\langle r \rangle_{\hat U}$ and $\langle r \rangle_{\hat U_0}$ (see Fig.~\ref{fig2} (c) and main text).
%The blue shaded region $\omega \lesssim 3J$ corresponds to the limit where the thermalization rate is slower and where $m>10$ is required to reach the PT distribution [see Fig.~\ref{fig3} (a) with $\omega = 2J$].
The other parameters are as in Fig.~\ref{fig2}.}
\label{fig_pd_kld}
\end{figure*}

%%%%%%%%%%%%%%%%%%%%%%%%%%%%%%%%%%%%%%%%%%%%%%%%%%%%%%%%%%%%%%%%%%%%%%%%%%%%%%%%%%%%%
%%%%%%%%%%%%%%%%%%%%%%%%%%%%%%%%%%%%%%%%%%%%%%%%%%%%%%%%%%%%%%%%%%%%%%%%%%%%%%%%%%%%%

\section{Quantum phases from supremacy signatures: A new order parameter}

Now that we have presented the different phases and discussed the computational complexity to sample from their dynamics, 
we analyse how quantum supremacy signatures, here taken to be the KLD to PT, can be directly used to probe the dynamical phase diagram of the driven Ising chain. 
The motivation is not only to provide an alternative order parameter, compared to the mean level spacing $\langle r \rangle$ and the population imbalance for example, but also to revisit these phase transitions from a computational complexity point of view.

\subsection{Phase diagram}

The protocol consists of initializing the system in a product state $\ket{\psi_0} = \otimes_i^L \ket{\uparrow}$, letting it evolve for $m$ driving cycles and compute its output probability distribution Pr$(p)$ [$p = p_m({\bf z})$].
We repeat the procedure for 100 disorder realizations and average the results.
In Fig.~\ref{fig_pd_kld} (a), we construct a ``phase diagram'' from the KLD of the output distributions to PT as a function of the disorder strength $W/J$ and the drive frequency $\omega/J$.
The solid black lines represent the parameter boundaries with $\langle r \rangle_{\hat U} = 0.51$ as in Fig.~\ref{fig2} for easier comparison. 
From the low-frequency and low-disorder enclosed region corresponding to $\langle r \rangle_{\hat U} > 0.51$, we see that the convergences of Pr$(p)$ towards the PT distribution at $m=10$ coincide with the convergence of the unitary $\hat U$ towards the COE.
The divergence from the PT distribution as $W \rightarrow 0$ is again a consequence of the non interacting nature of the model described by Eq.~\eqref{H_ising} with $W=0$.
%Within this region, we also observe a clear divergence from the PT distribution as $\omega \rightarrow 0$. 
%This is a consequence of slower thermalization rates [see Fig.~\ref{fig3} (a)] where larger evolution times (larger $m$) are needed to reach PT.
The PT is also not reached in the high-frequency enclosed region where $\hat U$ converges towards the GOE in the prethermalized regime.
%Those two limits are however different in nature. 
%For low frequencies, the state still reaches PT, albeit at a slower rate, while for high frequencies, the PT is never reached [see Fig.~\ref{fig3} (a)].
As discussed earlier, the computational complexity associated with this prethermalized regime has yet to be formally identified.

In Fig.~\ref{fig_pd_kld} (b) and (c), we show a cut along $\omega = 8J$ and $W = 4J$ respectively. We explicitly see the ability of the KLD to PT to capture the changes in $\langle r \rangle_{\hat U}$ and thus the phase transition from driven thermalized to MBL (panel b) and to the prethermalized regime (panel c). 
%In panel (c), the divergence from PT when $\omega \lesssim 3J$ and in the prethermalized regime is explicitly captured. 
Furthermore, it shows that the output probability distribution starts converging towards PT when the dynamics stop being accurately described by the time-averaged Hamiltonian $\hat H_0$.

%The success of recovering the phase diagram with the straightforward quantum supremacy signature {\it i.e.} KLD to PT suggests a more general claim that {\it any} suitable quantum supremacy signature can in principle be used to distinguish the different phases analyzed here. To further support this, we discuss the entanglement entropy as another possible option briefly in section~\ref{Sec:pt_inf_temp} and with more details in Appendix~\ref{App:EE}.

%%%%%%%%%%%%%%%%%%%%%%%%%%%%%%%%%%%%%%%%%%%%%%%%%%%%%%%%%%%%%%%%%%%%%%%%%%%%%%%%%%%%%
%%%%%%%%%%%%%%%%%%%%%%%%%%%%%%%%%%%%%%%%%%%%%%%%%%%%%%%%%%%%%%%%%%%%%%%%%%%%%%%%%%%%%

\subsection{Porter-Thomas distribution and infinite temperature thermalization}\label{Sec:pt_inf_temp}

So far, the discussions concerning the computational complexity of sampling from the driven thermalized phase mostly came from a complexity theory point of view (including Ref.~\cite{P1}) with the only intuition that the PT distribution is a signature of the dynamics with a capability to explore its entire Hilbert space. Here we provide additional physical intuitions by making a direct link between the PT distribution and thermalization to infinite temperature. 
This relation highlights the crucial role of the drive in providing the required energy to reach the PT distribution in our finite-size model.

%%%%%%%%%%%%%%%%%%%%%%%%%%%%%%%%%%%%%%%%%%%%%%%%%%%%%%%%%%%%%%%%%%%%%%%%%%%%%%%%%%%%%
%%%%%%%%%%%%%%%%%%%%%%%%%%%%%%%%%%%%%%%%%%%%%%%%%%%%%%%%%%%%%%%%%%%%%%%%%%%%%%%%%%%%%
In an isolated quantum system $\mathcal{S}_{\rm tot}$ described by a pure state $\ket{\Psi}$, the concept of thermalization is usually understood as the emergence of homogeneous statistical properties of the reduced density matrices $\hat \rho_\mathcal{S} = {\rm Tr}_{\mathcal{B}} \ket{\Psi}\bra{\Psi} \equiv \exp{(-\hat H_\mathcal{S}/k_BT_{\rm eff})}/\mathcal{Z}$.
Here, $\mathcal{S}$ denotes any ``small" subsystems with corresponding Hamiltonian $\hat H_\mathcal{S}$ of dimensions $N_\mathcal{S}$ and partition function $\mathcal{Z}$ such that $\mathcal{S}_{\rm tot} = \mathcal{S} + \mathcal{B}$ and $N_\mathcal{B} \gg N_\mathcal{B}$. 
The effective temperature $T_{\rm eff}$ is defined from the reduced density matrices and should become independent of frequency and of the subsystem choice as the full system reaches thermalization~\cite{2016_Alessio_AiP,pre_review}.  
By writing $\ket{\Psi} = \sum_i^{N_{\mathcal{S}}} \sum_j^{N_{\mathcal{B}}} c_{ij} \ket{i_\mathcal{S}} \ket{j_\mathcal{B}}$, where $\ket{j_\mathcal{S}}$ and $\ket{i_\mathcal{B}}$ are basis states of the subsystem and the bath respectively with $c_{ij} \in {\mathbb C}$, we obtain
\begin{equation} \label{Eq:pS}
\hat \rho_{\mathcal{S}} = \sum_{i,j = 1}^{N_{\mathcal{S}}} \left[ \sum_{k=1}^{N_{\mathcal{B}}} c_{ik} c^*_{jk} \right] \ket{i_\mathcal{S}}\bra{j_\mathcal{S}}.
\end{equation}
In the case where $N_{\mathcal{B}} \gg 1$ and the output probability distribution of $\ket{\Psi}$ follows the PT distribution given in Eq.~\eqref{Eq:PTD}, it is straightforward to show that $\bra{i_\mathcal{S}}\hat \rho_{\mathcal{S}}\ket{j_\mathcal{S}} = \sum_{k=1}^{N_{\mathcal{B}}} c_{ik} c^*_{jk} = \delta_{ij} / N_{\mathcal{S}}$ with $\delta_{ij}$ being the Kronecker delta function.
The vanishing off-diagonal elements come as a consequence that the amplitudes $c_{ik}$ underlying the PT distribution are random complex numbers subjected to the normalization constraint $\bra{\Psi}\Psi\rangle = 1$.
This result leads to an effective infinite temperature $T_{\rm eff} \rightarrow \infty$.
Furthermore, given that $\hat{\rho}_{\mathcal{S}} \sim \hat 1$ has to hold for all subsystems with $N_{\mathcal{B}} \gg 1$ to reach thermalization, $c_{ik}$ being random complex numbers is the only possible structure. 
In other words, a quantum state following PT corresponds to an infinite temperature thermalized state and vice versa.

Interestingly, this link between PT distribution and infinite temperature thermal phase suggests an alternative quantum supremacy signature that could be probed to observe the phase transitions. 
One could consider to use the difference between the entanglement entropy $S_e$ (see section \ref{Sec:MBL}) of all possible subsystems $\{\mathcal{S}\}$ and the corresponding maximal (infinite temperature) entropy $S_{\rm max} = \log_2(N_{\mathcal S})$ as an order parameter. More details concerning the entanglement entropy as a possible alternative route are given in Appendix~\ref{App:EE}.

%%%%%%%%%%%%%%%%%%%%%%%%%%%%%%%%%%%%%%%%%%%%%%%%%%%%%%%%%%%%%%%%%%%%%%%%%%%%%%%%%%%%%
%%%%%%%%%%%%%%%%%%%%%%%%%%%%%%%%%%%%%%%%%%%%%%%%%%%%%%%%%%%%%%%%%%%%%%%%%%%%%%%%%%%%%

\subsection{Drive-induced long-range multi-body interactions} \label{Sec:Magnus}

%We further discuss the key role of the drive in the Floquet formalism where we highlight the importance of the induced long-range multi-body interaction. 
More physical intuitions can be gained by highlighting the crucial role of the effective long-range multi-body interactions induced by the drive in facilitating a more efficient exploration of the Hilbert space. 
To see this, it is insightful to investigate the time-independent Floquet Hamiltonian $\hat{H}_F$.
For interacting systems, it is generally impossible to find an analytic form for $\hat H_F$ and one needs to resort to approximations. The most common approach is using the Magnus expansion to expand $\exp \left[-i H_F T\right]$ in a power series of $E_c/\omega$, where $E_c$ is a characteristic energy of the Hamiltonian $\hat H(t)$~\cite{mag_original,2015_Bukov_AiP,mag_review}.
%which, in our case, depends on $J, W, B_0$ and $\delta B$. 
%The latter can be the energy bandwidth or the variance of the eigenenergies of the undriven system. 
Doing so, one can write the Floquet Hamiltonian as $\hat{H}_{F}=\sum_{l=0}^{\infty} \hat{H}_{F}^{(l)}$, where the two first terms, for example, read
\begin{gather} \label{eq:ME}
    \hat H_F^{(0)} = \frac{1}{T}\int_{0}^{T}d\tau_1  \hat H(\tau_1) =\hat{H}_0, \\
    \hat H_F^{(1)} = \frac{1}{2iT}\int_{0}^{T}d\tau_1\int_{0}^{\tau_1}d\tau_2  [\hat H(\tau_1),\hat H(\tau_2)].
\end{gather}
The first term is the time average of $\hat{H}(t)$ which equals $\hat H_0$ and the next correction $H_F^{(2)}$ is computed explicitly in Appendix~\ref{App:MagExp}. In the limit of infinite frequencies $\omega \rightarrow\infty$, $\hat H_F$ converges to $\hat H_0$, while the series diverges in the low frequency regime $\omega < E_c$.

The crucial point is that higher-order contributions generally include multi-body and longer-range effective interactions. For example, $\hat H_F^{(2)}$ includes three-body interaction terms of the form $\hat H_i \sim \frac{J^2}{\omega^2}\delta B\hat \sigma^z_{i-1}\hat \sigma^x_{i} \hat \sigma^z_{i+1}$  [cf.~Eq.~\eqref{eq:hf2} in Appendix~\ref{App:MagExp}]. This tendency suggests that in the limit where the series diverges, the dynamics is governed by infinitely-long range multi-body interactions. 
In Fig.~\ref{fig2} (b) we show that as the driving frequency decreases, the level statistics of $\hat U$ starts converging toward the COE as the higher-order terms in the Magnus expansion start playing an increasingly important role. 
This is captured by comparing the statistics of the full evolution $\hat U$ with the statistics of the evolution $\hat U_0$ solely generated by the averaged Hamiltonian $\hat H_0$.
The same is observed in Fig.~\ref{fig_pd_kld} (a) where the output probability distribution Pr$(p)$ starts converging toward the PT distribution only once the evolution under $\hat H_0$ fails at approximating the exact evolution.  
This observation sheds light on the advantages of periodically driven systems to explore larger regions of their Hilbert space by highlighting the effectively enhanced connectivity of the model, which leads to faster growth of entanglement.
Note that two-body long-range interactions in 1D systems have been shown to slow down the entanglement entropy growth compared to nearest-neighbour interacting models~\cite{entropy_vary_range_prx2013}.
This results further highlights the importance of the multi-body nature of the induced interactions.  
In Appendix~\ref{App:AvsD}, we discuss how consequently driven analog quantum simulators should reach the quantum supremacy regime more rapidly compared to digital random quantum circuits with the same topology.

%%%%%%%%%%%%%%%%%%%%%%%%%%%%%%%%%%%%%%%%%%%%%%%%%%%%%%%%%%%%%%%%%%%%%%%%%%%%%%%%%%%%%
%%%%%%%%%%%%%%%%%%%%%%%%%%%%%%%%%%%%%%%%%%%%%%%%%%%%%%%%%%%%%%%%%%%%%%%%%%%%%%%%%%%%%

\section{Conclusion and outlooks}

In this work, we have described how key signatures of quantum supremacy in driven analog interacting systems can be directly used to probe its dynamical phase diagram. More precisely, we have focused on showing how the KLD between the distribution of the output state probabilities and the PT distribution can play the role of an order parameter capturing the transitions from the driven thermalized phase to the driven MBL phase and the Floquet prethermalized regime.
In this context, we have highlighted the clear connection between the effective multi-body long-range interactions generated by the drive and the capability of the system to reach the PT distribution. 
Compared to the commonly measured population imbalance~\cite{mbl_Rev,mbl_exp_2019_prx_bloch_2D,mbl_exp_2019_prl_bloch,mbl_exp_scq_chiaro2019direct} which captures the diffusion of an initial state in real space, the KLD to PT tells us about the diffusion over the Hilbert space. 
%Moreover, the output distribution is easier to access experimentally than the level statistics in driven systems for which no efficient measurement schemes are known so far.
Beside proposing an alternative order parameter for dynamical phase transitions, this work adopts a different angle more oriented towards the complexity theory. It represents a direct application of the very same protocol proposed to demonstrate quantum supremacy in generic driven many-body systems.
An interesting next step would be to analyse in more details the computational complexity aspects of the Floquet prethermalized regime.

%%%%%%%%%%%%%%%%%%%%%%%%%%%%%%%%%%%%%%%%%%%%%%%%%%%%%%%%%%%%%%%%%%%%%%%%%%%%%%%%%%%%%
%%%%%%%%%%%%%%%%%%%%%%%%%%%%%%%%%%%%%%%%%%%%%%%%%%%%%%%%%%%%%%%%%%%%%%%%%%%%%%%%%%%%%

\section{Acknowledgement}
This research is supported by the National Research Foundation, Prime Minister's Office, Singapore and the Ministry of Education, Singapore under the Research Centres of Excellence programme. It was also partially funded by Polisimulator project co-financed by Greece and the EU Regional Development Fund, Ninnat Dangniam is supported by the National Natural Science Foundation
of China (Grant No. 11875110).

%%%%%%%%%%%%%%%%%%%%%%%%%%%%%%%%%%%%%%%%%%%%%%%%%%%%%%%%%%%%%%%%%%%%%%%%%%%%%%%%%%%%%
%%%%%%%%%%%%%%%%%%%%%%%%%%%%%%%%%%%%%%%%%%%%%%%%%%%%%%%%%%%%%%%%%%%%%%%%%%%%%%%%%%%%%

\appendix

%%%%%%%%%%%%%%%%%%%%%%%%%%%%%%%%%%%%%%%%%%%%%%%%%%%%%%%%%%%%%%%%%%%%%%%%%%%%%%%%%%%%%
%%%%%%%%%%%%%%%%%%%%%%%%%%%%%%%%%%%%%%%%%%%%%%%%%%%%%%%%%%%%%%%%%%%%%%%%%%%%%%%%%%%%%

\section{MBL phase} \label{App:MBL}

We here provide a more detailed mathematical description of the MBL phase. We focus on demonstrating that the probability distribution $p_m({\bf z})$ of a quantum state evolving under its dynamics failed to satisfy the anti-concentration condition given in Eq.~\eqref{Eq:AC}. Finally, we discuss the feasibility to efficiently construct a classical sampler $\mathcal{C}$ capable of providing bitstrings from a distribution $q({\bf z})$ that is additively closed to $p_m({\bf z})$ as defined in Eq.~\eqref{add_err}.

%%%%%%%%%%%%%%%%%%%%%%%%%%%%%%%%%%%%%%%%%%%%%%%%%%%%%%%%%%%%%%%%%%%%%%%%%%%%%%%%%%%%%
%%%%%%%%%%%%%%%%%%%%%%%%%%%%%%%%%%%%%%%%%%%%%%%%%%%%%%%%%%%%%%%%%%%%%%%%%%%%%%%%%%%%%

\subsection{General framework}

The standard approach to describe the MBL phase is via a complete set of quasi-local integral of motions~\cite{mbl_tau_abanin,mbl_tau_huse,mbl_tau_imbre,mbl_Rev}, defined as $[\hat{\tau}^z_i, \hat H_{\rm MBL}] = 0$, where 
\begin{equation}\label{rela}
	\hat{\tau}^z_i = Z \hat{\sigma}^z_i + \sum_{m=1} e^{-m/\xi} \hat{O}^{(m)}_i.
\end{equation}   
Here, $Z \approx 1$, $\xi \ll 1$ is the characteristic localization length and $\hat{O}^{(m)}_i$ are generic operators describing interaction between up to $(2m+1)-$body over sites from $i-m$ to $i+m$.
A generic MBL Hamiltonian then adopts the form 
\begin{align}\label{mbl_h}
		\hat{H}_{\rm MBL} = & \sum_i h_i \hat{\tau}^z_i + \sum_{i < j} J_0 e^{-|i-j|/\xi'} \hat{\tau}^z_i\hat{\tau}^z_j \nonumber \\
		&+ \sum_{i < j <k} J_0 e^{-|i-k|/\xi'} \hat{\tau}^z_i\hat{\tau}^z_j \hat{\tau}^z_k + ...,
\end{align}
where $h_i \in [0,W]$ are on-site energies uniformly drawn from a random distribution, $J_0$ is an effective interaction strength and $\xi' \sim \xi$ is an additional localization length. 
%Various MBL features such as logarithmic growth of entanglement and dephasing can be understood with this qLIOM framework~\cite{mbl_tau_abanin,mbl_tau_huse}. 
For periodically driven systems, the standard approach is to cast the Floquet Hamiltonian $\hat{H}_F$ onto the form of Eq.~\eqref{mbl_h}~\cite{mbl_driven_2016_Abanin_AoP}. 

%%%%%%%%%%%%%%%%%%%%%%%%%%%%%%%%%%%%%%%%%%%%%%%%%%%%%%%%%%%%%%%%%%%%%%%%%%%%%%%%%%%%%
%%%%%%%%%%%%%%%%%%%%%%%%%%%%%%%%%%%%%%%%%%%%%%%%%%%%%%%%%%%%%%%%%%%%%%%%%%%%%%%%%%%%%

\subsection{Concentration of output distribution}

By inserting Eq.~(\ref{rela}) into Eq.~(\ref{mbl_h}), The MBL Hamiltonian can be expanded as a perturbation series
\begin{align}
	\hat{H}_{\rm MBL}  = \hat{H}_0 + \sum_{m=1}^{\infty}\sum_{i=1}^L \epsilon^m \hat{V}^{(m)}_i,
\end{align} 
where $\epsilon = e^{-1/\xi} \ll 1$ (we took $\xi' = \xi$ for simplicity), $\hat{H}_0=\sum_i h_i Z\hat{\sigma}^z_i$ is the non-interacting part  and $\hat{V}^{(m)}_i$ are local perturbations acting on at most sites from $i - m$ to $i+m$. 
At first order perturbation ($m=1$), $\hat{V}^{(1)}_i$ can flip at max 3 {\it neighbouring} spins
and can thus only mix an unperturbed product state of $L$ spins $\ket{{\bf z}_0}$ with up to $4L$ different basis states (neglecting the boundary corrections). 
The same logic applies at second order perturbation where this time $\ket{{\bf z}_0}$ becomes a linear superposition of up to $16L$ additional basis states. 
As the order of perturbation increases, a polynomial number (poly($L$)) of additional basis states have to be taken into account.
This is a consequence of the local nature of each perturbation term. In contrast, extended interactions, as it would be the case in the thermal phase, would lead to an exponential increase of additional basis states. 
Assuming that an initial product state evolves deep inside the MBL phase where only few orders of perturbation is required, the output state would only have a polynomial number of basis state populated.
Since the anti-concentration condition in Eq.~(\ref{Eq:AC}) requires most of the $2^L$ basis states to be populated, the output distribution in the MBL phase can not satisfy it.

%%%%%%%%%%%%%%%%%%%%%%%%%%%%%%%%%%%%%%%%%%%%%%%%%%%%%%%%%%%%%%%%%%%%%%%%%%%%%%%%%%%%%
%%%%%%%%%%%%%%%%%%%%%%%%%%%%%%%%%%%%%%%%%%%%%%%%%%%%%%%%%%%%%%%%%%%%%%%%%%%%%%%%%%%%%

\begin{figure*}
    \includegraphics[width= 1\textwidth]{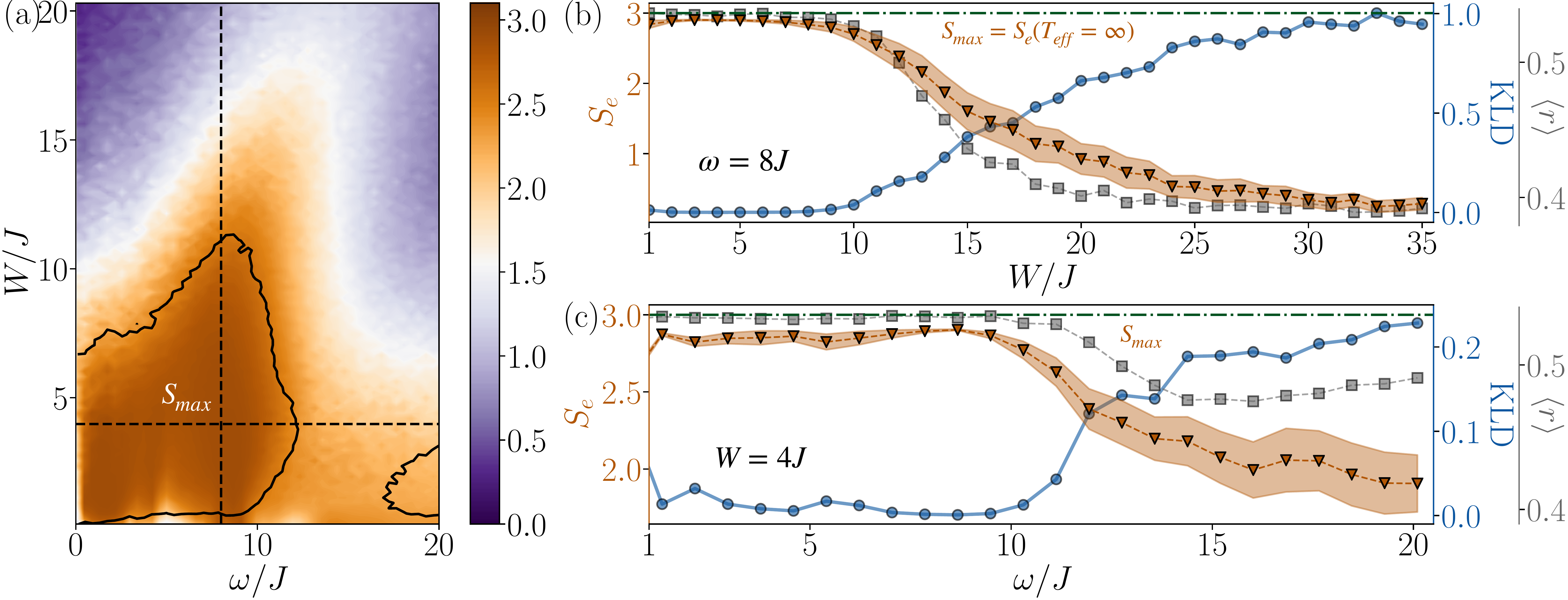}
	\caption{\textbf{Probing phase diagram with entanglement entropy.} (a) The average entanglement entropy over six different subsystems, each containing three sites chosen randomly, is used as an alternative quantum supremacy signature to revisit the phase diagram. The vertical and horizontal cuts show the average entanglement entropy as a function of (b) $W/J$ and (c) $\omega/J$. The shaded orange region indicates one standard deviation. The horizontal dashed lines represent the maximal entropy at the infinite temperature $S_{\rm max}=S_e(T_{\rm eff}=\infty)$. The KLD to PT $\text{KLD}(\text{Pr}(p)\parallel\text{PT}(p))$ (blue circles) and the average level spacing $\langle r \rangle_{\hat U}$ (grey squares) are also plotted for comparison. All parameters are identical to those in Fig.~\ref{fig_pd_kld}. The low variance in the thermalized phase indicates the homogeneity.} \label{Fig:EE}
\end{figure*}

\subsection{Complexity of sampling task}
Previous works have demonstrated the ability of MPS states to efficiently approximate all eigenstates of any MBL Hamiltonians~\cite{mbl_mps_jens_PhysRevLett,mbl_Rev}. 
In what follows we denote an exact eigenstate of $\hat H_{\rm MBL}$ with eigenenergy $E_k$ as $\ket{\varphi_k}$ and the associated MPS approximation as $\ket{{\rm MPS}_k}$.
We define the error made by the approximation, $\nu \ll 1$, as
\begin{align}\label{mps_mbl}
    \ket{{\rm MPS}_k}& = (1-\nu)\ket{\varphi_k} + \sqrt{2\nu-\nu^2} \sum_{k'\neq k} a_{kk'} \ket{\varphi_{k'}} ,\nonumber \\
    & \approx \ket{\varphi_k} + \sqrt{2\nu} \sum_{k'\neq k} a_{kk'} \ket{\varphi_{k'}},
\end{align}
where $\sum_{k'\neq k}a_{kk'}$ captures the overlap over all other eigenstates $\ket{\varphi_{k'}}$ with $\sum_{k'\neq k} |a_{kk'}|^2 = 1$. 
The mean energy associated with $\ket{{\rm MPS}_k}$ can be written as
\begin{align}
    \tilde{E}_k & = \bra{{\rm MPS}_k} \hat{H}_{\rm MBL} \ket{{\rm MPS}_k},\nonumber \\
    & \approx E_k + 2\nu \sum_{k'\neq k} |a_{kk'}|^2 E_{k'}, \nonumber \\
    & \equiv E_k + 2\nu \Lambda_{k},
\end{align}
where $\Lambda_{k'}$ represents the average over the entire energy spectrum following the ``error distribution'' $\sum_{k'\neq k} |a_{kk'}|^2$.
The bond dimension of the MPS states required to maintain a fix level of precision $\nu$ have been shown to scale polynomially with the system size $L$~\cite{mbl_mps_jens_PhysRevLett,mbl_Rev}; the exact scaling depends on the specifics of $\hat H_{\rm MBL}$.
The approximated output probability distribution $q_t(\textbf{z})$, after evolving from the initial state $\ket{{\bf z}_0}$ for a time $t$, then reads
\begin{widetext}
\begin{align} \label{Eq:qtMBL}
 q_t(\textbf{z}) & = \left| \bra{\textbf{z}}\left( \sum_k e^{-i\tilde E_k t }  \ket{\rm{MPS}_k}\bra{\rm{MPS}_k}\right)\ket{\textbf{z}_0} \right|^2 , \\
  & \approx \left| \bra{\textbf{z}} \sum_k e^{-i(E_k + 2\nu\Lambda_k)t} \left( \ket{\varphi_k}\bra{\varphi_k} + \sqrt{2\nu} \sum_{k'\neq k} \left[a_{kk'}\ket{\varphi_{k'}}\bra{\varphi_k}+ {\rm H.c.} \right]\right)\ket{\textbf{z}_0} \right|^2 , \nonumber \\
  & \approx \left| \bra{\textbf{z}} \sum_k e^{-i E_k t} \left( [1 - 2i\nu\Lambda_kt]\ket{\varphi_k}\bra{\varphi_k} + \sqrt{2\nu} \sum_{k'\neq k} \left[a_{kk'}\ket{\varphi_{k'}}\bra{\varphi_k}+ {\rm H.c.} \right]\right)\ket{\textbf{z}_0} \right|^2 , \nonumber \\
 & \approx p(\textbf{z}) + \left[\sqrt{2\nu} \sum_{k,k'} e^{-i(E_k-E_{k'})t} \langle \textbf{z} |\hat{P}_k|\textbf{z}_0\rangle \bra{\textbf{z}_0} \hat{M}_{k'}\ket{\textbf{z}}
 + 2i\nu t \sum_{k,k'} e^{-i(E_k-E_{k'})t} \Lambda_{k'}\langle\textbf{z} |\hat{P}_k|\textbf{z}_0\rangle \langle\textbf{z}_0 |\hat{P}_{k'}|\textbf{z}\rangle  + c.c. \right]. \nonumber
\end{align}
\end{widetext}
Here $\hat{P}_k =  \ket{\varphi_k}\bra{\varphi_k}$, $\hat{M}_k = \sum_{k'\neq k}a_{kk'}\ket{\varphi_{k'}}\bra{\varphi_k}+ {\rm H.c.}$ and the exact distribution is given by $p_t(\textbf{z})=\left| \sum_k e^{-iE_kt} \bra{\textbf{z}}\hat{P}_k\ket{\textbf{z}_0}\right|^2$.
In the above expression, we keep only the leading order in $\nu$ and $\nu t$ and we considered the limit $\nu\Lambda_kt \ll 1$.

In the short time limit where the last term in the last line of Eq.~\eqref{Eq:qtMBL} can be neglected, the error made in approximating $p_t(\textbf{z})$ is proportional to $\sqrt{2\nu} \sum_{k} e^{-iE_kt} \langle \textbf{z} |\hat{P}_k|\textbf{z}_0\rangle$.
Following the discussion in the previous subsection, we know that for a dynamics deep in the MBL phase, only a small fraction [poly$(L)/2^L$] of $\langle \textbf{z} |\hat{P}_k|\textbf{z}_0\rangle$ will be nonzero and at max proportional to $e^{-1/\xi}$.
Therefore, the additive error condition in Eq.~(\ref{add_err}), i.e.~$\sum_{\textbf{z}}|p_t(\textbf{z}) - q_t(\textbf{z})| \leq \beta_0$, is expected to be reachable with arbitrary small $\beta_0$ for bond dimensions of the MPS states that scale polynomially with the system size. 
For longer evolution times, the term proportional to $\nu t$ will start to become increasingly important.
It will inevitably reach a point where any small initial mistake $2\nu\Lambda_k$ made estimating the eigenenergies accumulates, leading to a large error in the phases.
While any approximation method should fail in this very large time limit, we note that the same is bounded to happen for a quantum simulator; any small amount of dephasing or decay will inevitably accumulate beyond the required accuracy in the large time limit. 

%%%%%%%%%%%%%%%%%%%%%%%%%%%%%%%%%%%%%%%%%%%%%%%%%%%%%%%%%%%%%%%%%%%%%%%%%%%%%%%%%%%%%

%%%%%%%%%%%%%%%%%%%%%%%%%%%%%%%%%%%%%%%%%%%%%%%%%%%%%%%%%%%%%%%%%%%%%%%%%%%%%%%%%%%%%

\section{Entanglement entropy} \label{App:EE}

As discussed in section \ref{Sec:pt_inf_temp}, dividing a closed system into two parts, ${\mathcal{S}}$ and ${\mathcal{B}}$ of size $N_{\mathcal{S}}$ and $N_{\mathcal{B}}$ respectively, and having a reduced density matrix $\hat \rho_{\mathcal{S}} = \hat 1 / N_{\mathcal{S}}$ for all possible subsystems $\mathcal{S}$ corresponds to an infinite-temperature thermalized state. Here the reduced density matrix is $\hat \rho_{\mathcal{S}} = {\rm Tr}_{\mathcal{B}}\ket\Psi\bra\Psi$ where $\ket\Psi$ is the state of the full closed system.
In that case, the output probability distribution of the quantum state $\ket\Psi$ follows the PT distribution and consequently fulfil the anti-concentration condition given in Eq.~\eqref{Eq:AC}.
This means that an alternative route to demonstrate the anti-concentration condition could be to measure the reduced density matrix of all subsystems $\mathcal{S}$ and show that $\hat \rho_{\mathcal{S}} = \hat 1 / N_{\mathcal{S}}$ homogeneously over the whole system. We note that entanglement entropy of undriven systems has been experimentally measured in different quantum platforms~\cite{mbl_Rev,2016_Alessio_AiP,mbl_exp_scq_chiaro2019direct,ee_lukin2015}.

In Fig.~\ref{Fig:EE}, we plot the entanglement entropy $S_e = {\rm Tr}\{\hat\rho_{\mathcal{S}}\log\hat\rho_{\mathcal{S}}\}$ as the function of $\omega$ and $W$ averaged over multiple subsystems. 
We choose six different subsystems consisting of three sites randomly selected and apply the same procedure as in the main text.
%for the same parameters as in Figs.~\ref{fig2} and \ref{fig_pd_kld} (b)-(c).  
We see that the system approaches the maximal entropy associated with the infinite temperature state, $S_{\rm max} = \log_2(N_{\mathcal{S}})$ when the output probability follows PT and the statistics of $\hat U$ converges toward the COE. The homogeneity of the thermalized phase is indicated by the low variance of $S_e$ over the different subsystems. 

%%%%%%%%%%%%%%%%%%%%%%%%%%%%%%%%%%%%%%%%%%%%%%%%%%%%%%%%%%%%%%%%%%%%%%%%%%%%%%%%%%%%%
%%%%%%%%%%%%%%%%%%%%%%%%%%%%%%%%%%%%%%%%%%%%%%%%%%%%%%%%%%%%%%%%%%%%%%%%%%%%%%%%%%%%%

%%%%%%%%%%%%%%%%%%%%%%%%%%%%%%%%%%%%%%%%%%%%%%%%%%%%%%%%%%%%%%%%%%%%%%%%%%%%%%%%%%%%%
%%%%%%%%%%%%%%%%%%%%%%%%%%%%%%%%%%%%%%%%%%%%%%%%%%%%%%%%%%%%%%%%%%%%%%%%%%%%%%%%%%%%%

%%%%%%%%%%%%%%%%%%%%%%%%%%%%%%%%%%%%%%%%%%%%%%%%%%%%%%%%%%%%%%%%%%%%%%%%%%%%%%%%%%%%%
%%%%%%%%%%%%%%%%%%%%%%%%%%%%%%%%%%%%%%%%%%%%%%%%%%%%%%%%%%%%%%%%%%%%%%%%%%%%%%%%%%%%%

\section{Magnus expansion of $\hat H(t)$.} \label{App:MagExp}
We here explicitly compute the Magnus expansion up to the second order correction. The first term reads
\begin{align}
\hat H_F^{(1)} &= \frac{1}{2iT}\int_{t_0}^{T+t_0}d\tau_1\int_{t_0}^{\tau_1}d\tau_2  \left[ \hat H(\tau_1),\hat H(\tau_2)\right] \\
&= \frac{1}{2iT}\int_{t_0}^{T+t_0}d\tau_1\int_{t_0}^{\tau_1}d\tau_2 \left[\hat{H}_{0},\hat{H}_d(\tau_2)-\hat{H}_d(\tau_1)\right] \nonumber \\
& \hspace{-1cm}= \frac{2i\delta B \sin{(\omega t_0})}{\omega} \left[ \sum_{j=1}^{L} h_j \hat{\sigma}^{y}_j + J \sum_{j=1}^{L-1}\left(\hat{\sigma}^{y}_j \hat{\sigma}^{z}_{j+1}+\hat{\sigma}^{z}_{j} \hat{\sigma}^{y}_{j+1}\right) \right]. \nonumber
\end{align}
The choice of $t_0 = 0$ is arbitrary and can lead to different forms of the Floquet Hamiltonian $\hat H_F$. Consistently with the main text, we fix $t_0 = 0$, leading to $\hat H_F^{(1)} = 0$.
The second order correction term adopts the form
\begin{widetext}
\begin{align}
\hat{H}_F^{(2)} = & -\frac{1}{6T}\int_{t_0}^{T+t_0}d\tau_1\int_{t_0}^{\tau_1}d\tau_2\int_{t_0}^{\tau_2}d\tau_3  %\nonumber \\ &
\times \left(\left[\hat{H}(\tau_1),\left[\hat{H}(\tau_2),\hat{H}(\tau_3)\right]\right]+(1\Leftrightarrow 3)\right), \nonumber
\\
= & \frac{-4\delta B }{\omega^2} \left\{  \sum_{j=1}^L h_j^2 \hat{\sigma}^{x}_j + 2J \sum_{j=1}^{L-1} \left(h_j \hat{\sigma}^{x}_j \hat{\sigma}^{z}_{j+1} + h_{j+1} \hat{\sigma}^{z}_j \hat{\sigma}^{x}_{j+1}\right)  %\right. \nonumber \\ & +  \left. 
+ J^2 \sum_{j=1}^{L-1} \left(\hat{\sigma}^{x}_j  + \hat{\sigma}^{x}_{j+1}\right) + 2 J^2 \sum_{j=1}^{L-2} \left(\hat{\sigma}^{z}_j \hat{\sigma}^{x}_{j+1} \hat{\sigma}^{z}_{j+2}\right) \right\} \nonumber \\
& + \frac{4B_0\delta B}{\omega ^2} \left\{  \sum_{j=1}^L h_j\hat{\sigma}^{z}_j  + 2J \sum_{j=1}^{L-1} \left(\hat{\sigma}^{z}_j \hat{\sigma}^{z}_{j+1} -\hat{\sigma}^{y}_j \hat{\sigma}^{y}_{j+1}\right) \right\}.
\label{eq:hf2}
\end{align}
\end{widetext}
%Note that $\hat{\sigma}^{z}_j \hat{\sigma}^{x}_{j+1} \hat{\sigma}^{z}_{j+2}$ in the last line is the three-body interaction term which is also mentioned in the main text. The many-body long range interaction terms are common in the higher order corrections. Lastly, the higher order correction terms scale as $\left(\frac{E_c}{\omega}\right)^n$ where $E_c$ is the characteristic energy of the system.
As discussed in the main text, higher-order correction includes longer-range interactions involving increasing number of spins with coupling strength that scale proportionally to $E_c/\omega$. Multiple works have been dedicated to the study of the Magnus expansion convergence in driven systems by estimating the relevant energy scale $E_c$~\cite{mag_converge1,mag_converge2}.

%%%%%%%%%%%%%%%%%%%%%%%%%%%%%%%%%%%%%%%%%%%%%%%%%%%%%%%%%%%%%%%%%%%%%%%%%%%%%%%%%%%%%
%%%%%%%%%%%%%%%%%%%%%%%%%%%%%%%%%%%%%%%%%%%%%%%%%%%%%%%%%%%%%%%%%%%%%%%%%%%%%%%%%%%%%

\begin{figure*}
\includegraphics[width=1.0\textwidth]{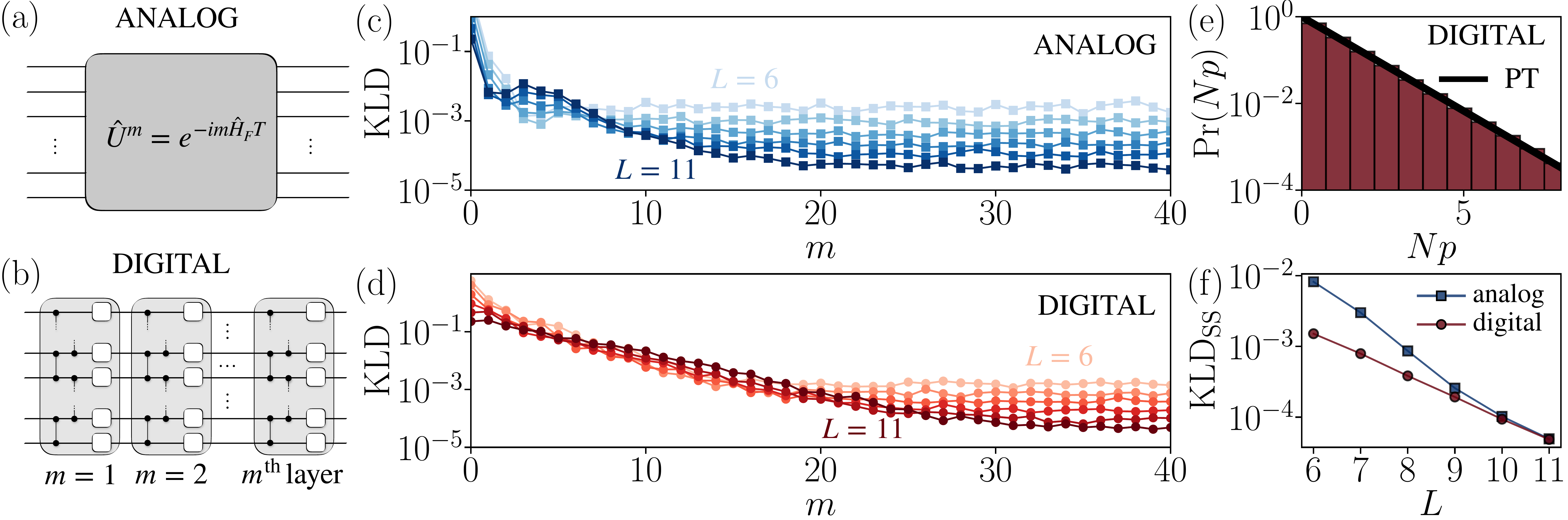}
\caption{\textbf{Benchmarking quantum supremacy with analog and digital approaches:} $\textbf{(a)}$ a circuit diagram of the analog driven quantum Ising chain with $m$ driving cycles. $\textbf{(b)}$ a circuit diagram of random quantum circuits containing $m$ layers of random single qubits in the set $\{\sqrt{X},\sqrt{Y},T\}$ and controlled-Z gates. $\textbf{(c)}$ ($\textbf{(d)}$)  $\text{KL}(\text{Pr}(p)\parallel\text{PT}(p))$ as a function of $m$ and the size of the system $L$ for the analog (digital) approach. $\textbf{(e)}$ The output distribution, weighted by $N=2^L$, in the analog case with $L=11$ and $m=40$. The yellow line is the exact PT distribution. $\textbf{(f)}$ KL divergence at the long time limit of the analog and the digital cases as a function of $L$. ($W=3J$, $B_0=-\delta B=1.25J$, $\omega=8J$ and 500 disorder realizations.) }
\label{fig_appx}
\end{figure*}

\section{Comparison between analog and digital approaches in reaching quantum supremacy} \label{App:AvsD}

In this appendix, we compare the time it takes for a quantum state to evolve toward the PT distribution under the dynamics of the driven analog Ising spin chain and the 1D digital random quantum circuits~\cite{2018_hartmut_natphy}.
To keep the comparison fair, we consider the same interaction topology, i.e.~a 1D system with nearest-neighbour couplings and open boundary conditions.
We show that the effective long-range multi-body interactions induced by the drive allow the analog quantum system to converge considerably faster in the regime where quantum supremacy can be demonstrated.

The model we use to simulate the random quantum circuits is based on Ref.\cite{2018_hartmut_natphy} and consists of $m$ layers of gates acting on $L$ qubits, as depicted in Fig.~\ref{fig_appx}(b). 
Each layer consists of a series of single-qubit gates followed by a series of controlled-Z gates. 
Moreover, a layer of Hadamard gates is added once at the very beginning of the circuit. 
Each single-qubit gates are chosen randomly from the set $\{\sqrt{X},\sqrt{Y},T\}$, where $\sqrt{X}$ ($\sqrt{Y}$) represents a $\pi/2$ rotation around the $X$ ($Y$) axis of the Bloch sphere and $T$ is a non-Clifford gate represented by the diagonal matrix $\{1,e^{i\pi/4}\}$. 
As in the analog case, we measure the output distribution ${\rm Pr}(p_m({\bf z}))$ averaged over $D$ distinct realizations of this model circuit. 
In order to compare the evolution time, we set the driving period of the analog Ising chain to be approximately equal to the duration of a single layer of gates ($m=1$) of the digital circuit. Since single-qubit gates can usually be implemented much faster compared to two-qubit gates, we assume that each layer takes the same time as a controlled-Z gate from a standard $J\sigma^z_i\sigma^z_{i+1}$ coupling, i.e.~$t_m \sim \pi/4J$. This results in a driving frequency $\omega=8J$. 

In Fig.~\ref{fig_appx}(c)-(d), we plot the $\text{KLD}(\text{Pr}(p)\parallel\text{PT}(p))$ with $p=p_m({\bf z})$ for the driven Ising chain and the digital random circuits, respectively, as a function of the driving cycle $m$ for different system sizes $L$. The long time limit is shown in Fig.~\ref{fig_appx} (e). 
We observe that, despite having the same topology, the driven analog system converges to the PT distribution considerably faster as a results of the enhanced connectivity induced by the drive. Given the limited coherence times in the current state-of-the-art quantum platforms, such gain makes the analog approaches more favorable as testbeds for quantum supremacy. Moreover, the analog approach does not require precise local control to implement the gates.

%%%%%%%%%%%%%%%%%%%%%%%%%%%%%%%%%%%%%%%%%%%%%%%%%%%%%%%%%%%%%%%%%%%%%%%%%%%%%%%%%%%%%
%%%%%%%%%%%%%%%%%%%%%%%%%%%%%%%%%%%%%%%%%%%%%%%%%%%%%%%%%%%%%%%%%%%%%%%%%%%%%%%%%%%%%
\bibliography{ref}

%%%%%%%%%%%%%%%%%%%%%%%%%%%%%%%%%%%%%%%%%%%%%%%%%%%%%%%%%%%%%%%%%%%%%%%%%%%%%%%%%%%%%
%%%%%%%%%%%%%%%%%%%%%%%%%%%%%%%%%%%%%%%%%%%%%%%%%%%%%%%%%%%%%%%%%%%%%%%%%%%%%%%%%%%%%

\end{document}